\documentclass[12pt]{article}
\usepackage{epsfig}
\usepackage{graphicx}
\usepackage{color}

\def\Title#1{\begin{center} {\Large #1 } \end{center}}

\newenvironment{Abstract}{\begin{quotation}  }{\end{quotation}}
\newenvironment{Presented}{\begin{quotation} \begin{center} 
             PRESENTED AT\end{center}\bigskip 
      \begin{center}\begin{large}}{\end{large}\end{center} \end{quotation}}

\def\beq{\begin{equation}}
\def\eeq#1{\label{#1}\end{equation}}
\def\eeqn{\end{equation}}

\def\beqa{\begin{eqnarray}}
\def\eeqa#1{\label{#1}\end{eqnarray}}
\def\eeqan{\end{eqnarray}}

\let\bar=\overbar

\def\Dslash{\not{\hbox{\kern-4pt $D$}}}
\def\dslash{\not{\hbox{\kern-2pt $\del$}}}

\def\msb{{\bar{\ssstyle M \kern -1pt S}}}

\newcommand{\pipi}{\ensuremath{B \rightarrow \pi \pi}}
\newcommand{\pippim}{\ensuremath{B^{0} \rightarrow \pi^{+} \pi^{-}}}

\newcommand{\aonepi}{\ensuremath{B^{0} \rightarrow a_{1}(1260)^{\pm} \pi^{\mp}}}

\newcommand{\pip}{\ensuremath{\pi^{+}}}
\newcommand{\pim}{\ensuremath{\pi^{-}}}

\newcommand{\Bz}{\ensuremath{B^{0}}}
\newcommand{\Bzb}{\ensuremath{\bar{B}^{0}}}
\newcommand{\YFS}{\ensuremath{\Upsilon(4S)}}

\newcommand{\Dt}{\ensuremath{\Delta t}}

\newcommand{\Acp}{\ensuremath{{A}_{CP}}}

\newcommand{\Scp}{\ensuremath{{S}_{CP}}}

\newcommand{\phitwo}{\ensuremath{\phi_{2}}}

\begin{document}
\Title{Results on $\phitwo$ from $e^+e^-$ colliders}
\bigskip\bigskip

\begin{raggedright}  
{\it Pit Vanhoefer\index{Vanhoefer, P.}\\
Max-Plank-Institut f\"ur Physik\\
 F\"oringer Ring 6\\
  M\"unchen 80805 GERMANY}
\bigskip\bigskip
\end{raggedright}


\begin{Abstract}
We present a summary of measurements sensitive to the CKM angle \phitwo\ ($\alpha$), performed by the BaBar and the Belle experiments which both collect $B\bar{B}$ pairs produced at the \YFS\ resonance in asymmetric $e^+e^-$ collisions. We discuss the decays $B \rightarrow \pi \pi, \rho\rho, a_1^{\pm}\pi^{\mp}$ and $(\rho\pi)^0$. The $CP$ asymmetries, branching fractions and polarizations obtained are used to constrain $\phi_2$.
\end{Abstract}
\vfill
\begin{Presented}
FPCP2013, the 7th International Workshop on the CKM Unitarity Triangle\\
Buzios Brazil
\end{Presented}
\vfill
\centering
 \includegraphics[height=!,width=0.15\columnwidth]{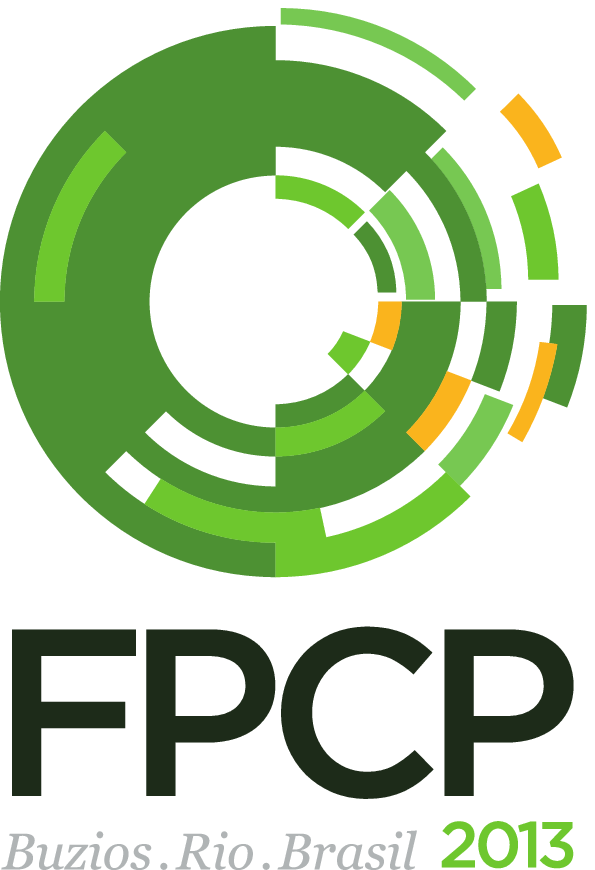}
\def\thefootnote{\fnsymbol{footnote}}
\setcounter{footnote}{0}

\section{Introduction}
\begin{figure}[htb]
  \centering
 \includegraphics[height=90pt,width=!]{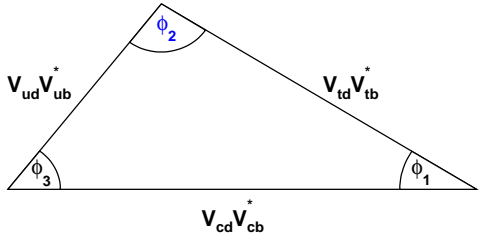} 
  \caption{Sketch of the unitarity triangle for $B$ decays with the definition of the internal angles.}
  \label{p_phi2a}
\end{figure}

A major precision test of the Standard Model (SM) is to validate the Cabibbo-Kobayashi-Maskawa (CKM) mechanism for violation of the combined charge-parity ($CP$) symmetry~\cite{C,KM}. This is one of the main purposes of the Belle and BaBar experiments, which were operating at the KEKB and PEPII $B$-factories, respectively. Both experiments contributed significantly to probing the CKM scheme, constraining the unitarity triangle for $B$ mesons to its current precision. Any deviation from unitarity would be a clear hint for physics beyond the SM. These proceedings give a summary of the experimental status of measurements of the CKM angle \phitwo\ ($\alpha$) defined from the CKM matrix elements $V_{ij}$ as $\phitwo \equiv \arg(-V_{td}V^{*}_{tb})/(V_{ud}V^{*}_{ub})$, shown in Fig~\ref{p_phi2a}. Both experiments finished their data taking; BaBar collected $465\times 10^6$ $B\bar{B}$ pairs and Belle $772\times 10^6$ $B\bar{B}$. 

In the quasi-two-body approach, the CKM angles can be determined by measuring the time-dependent asymmetry between \Bz\ and \Bzb\ decays into a common $CP$ eigenstate~\cite{CP}. In the decay sequence, $\YFS \rightarrow B_{CP}B_{\rm Tag} \rightarrow f_{CP}f_{\rm Tag}$, where one of the $B$ mesons ($B_{CP}$) decays into a $CP$ eigenstate  $f_{CP}$ at a time $t_{CP}$ and the other ($B_{\rm Tag}$) decays into a flavor specific final state $f_{\rm Tag}$ at a time $t_{\rm Tag}$, the time-dependent decay rate is given by
\begin{equation}
  {P}(\Delta t, q) = \frac{e^{-|\Dt|/\tau_{B^0}}}{4\tau_{B^0}} \bigg[ 1+q(\Acp\cos\Delta m_d \Dt + \Scp\sin\Delta m_d \Dt) \bigg],
\label{eq1}
\end{equation}
where $\Dt = t_{CP}- t_{\rm Tag}$, $\Delta m_d$ is the mass difference between the mass eigenstates  $B_{H}$ and $B_{L}$ and $q = \pm 1$ for $B_{\rm Tag} = \Bz (\Bzb)$. The $CP$ asymmetry is given by 
\begin{equation}
\frac{N(\bar{B}\to f_{CP}, \Dt) - N(B\to f_{CP}, \Dt)}{N(\bar{B}\to f_{CP}, \Dt) + N(B\to f_{CP}, \Dt)},
\label{eq_asym}
\end{equation}
where $ N(B(\bar{B})\to f_{CP})$ is the number of events of a $B(\bar{B})$ decaying to $f_{CP}$.
 The parameters \Acp\ and \Scp\ describe direct and mixing-induced $CP$ violation, respectively~\footnote{There exists an alternative notation: $C_{CP} = -\Acp$.}.

\section{Principle Of Determining \phitwo}
\begin{figure}[htb]
\begin{minipage}[]{0.49\columnwidth} 
  \centering
 \includegraphics[height=120pt,width=!]{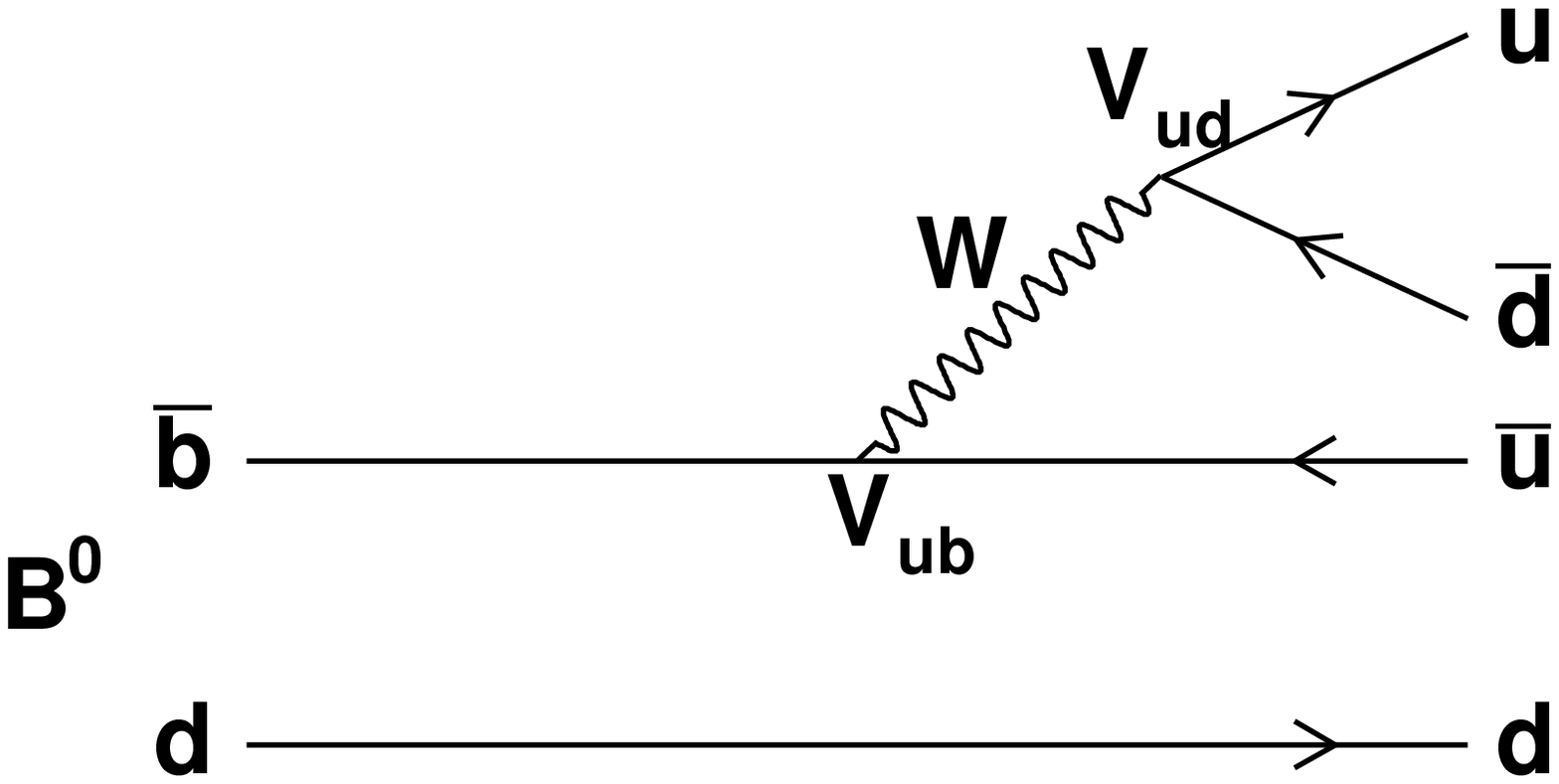} 
\put(-200, 100){a)}
\end{minipage}
\begin{minipage}[]{0.49\columnwidth} 
  \centering
 \includegraphics[height=120pt,width=!]{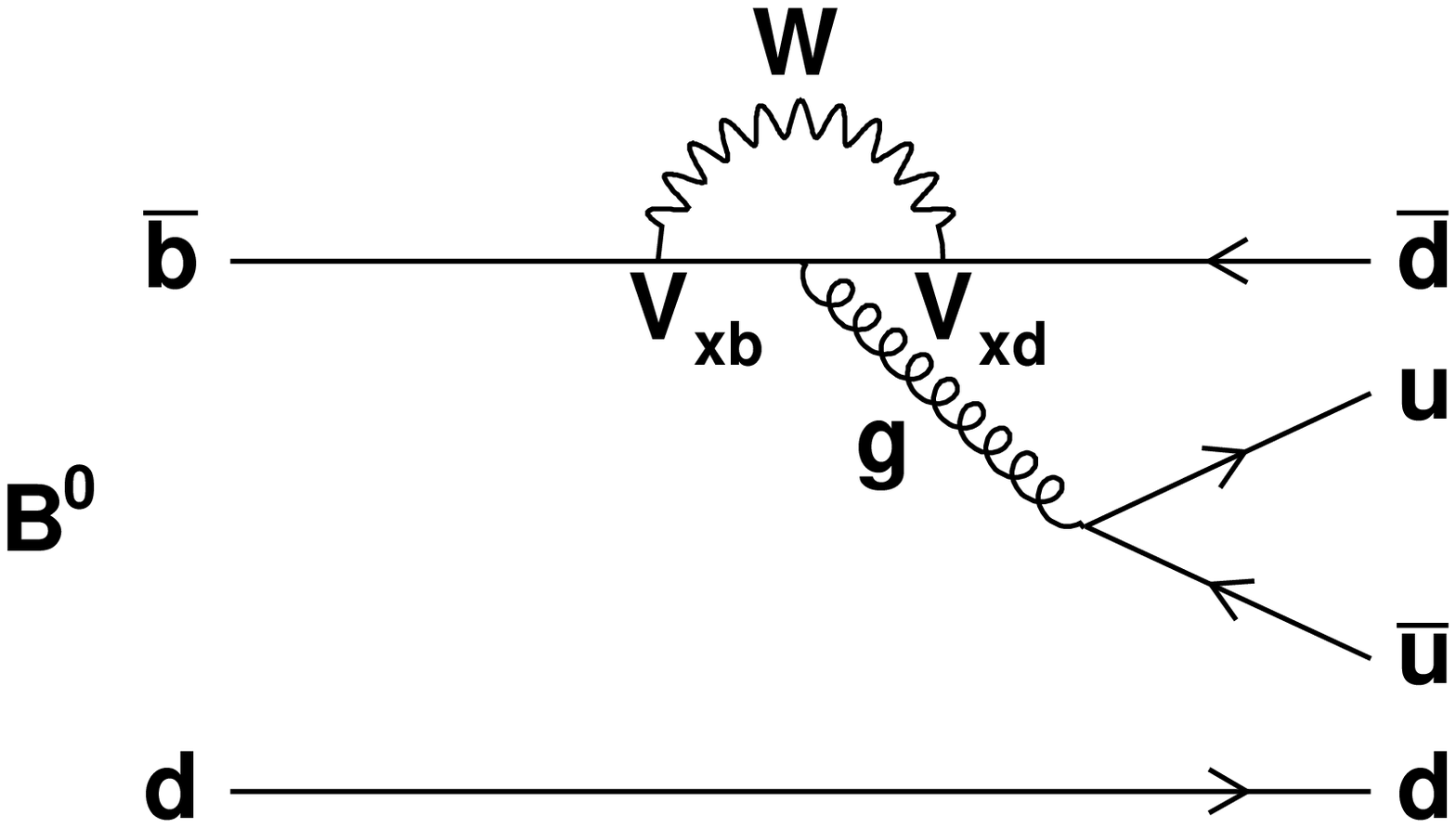} 
\put(-200, 100){b)}
\end{minipage}
  \caption{a) Tree and b) penguin (second order) feynman diagrams of $B^0\to h^+h^-$.}
  \label{p_feyn}
\end{figure}

Decays proceeding via $b \rightarrow u \bar{u} d$ quark transitions such as $B \rightarrow \pi\pi$, $\rho\pi$, $\rho\rho$ and $a_1\pi$, (see also Fig.~\ref{p_feyn}) are sensitive to \phitwo\ via mixing induced $CP$ violation. At tree level we expect $\Acp=0$ and $\Scp=\sin2\phitwo$. However, possible penguin contributions can give rise to direct $CP$ violation, $\Acp\neq 0$ and also pollute the measurement of \phitwo, $\Scp=\sqrt{1-\Acp^{2}}\sin(2\phitwo^{eff})$, where the observed $\phitwo^{eff} \equiv \phitwo - \Delta \phitwo$ is shifted by $\Delta \phi_2$ due to different weak and strong phases from the additional contributions.

Despite this, it is possible to determine $\Delta \phitwo$ in $\Bz \rightarrow h^{+} h^{-}$ with a $SU(2)$ isospin analysis by considering the set of three $B \rightarrow hh$ decays where the final state $hh$ consists either of two pions or two longitudinally polarized $\rho$ mesons, each related via isospin symmetry~\cite{iso}. The $B \rightarrow hh$ amplitudes obey the triangle relations (Fig.~\ref{p_phi2})
\begin{equation}
  A_{+0} = \frac{1}{\sqrt{2}}A_{+-} + A_{00}, \;\;\;\; \bar{A}_{-0} = \frac{1}{\sqrt{2}}\bar{A}_{+-} + \bar{A}_{00},
  \label{eq_iso}
\end{equation}
where the subscript denotes the charges of the final state mesons. In the limit of neglecting electroweak penguins and isospin breaking effects $B^{\pm} \rightarrow h^{\pm} h^{0}$ is a pure first-order mode, thus these triangles share the same base, $A_{+0}=\bar{A}_{-0}$. $\Delta \phitwo$ can then be determined from the difference between the two triangles. Because of the four possible orientations of the two isospin triangles, this method has an inherent 8-fold discrete ambiguity in the determination of \phitwo.
There are further methods to extract $\phi_2$ using the $SU(3)$ symmetry~\cite{CKMangles}.

\begin{figure}[htb]
  \centering
 \includegraphics[height=120pt,width=!]{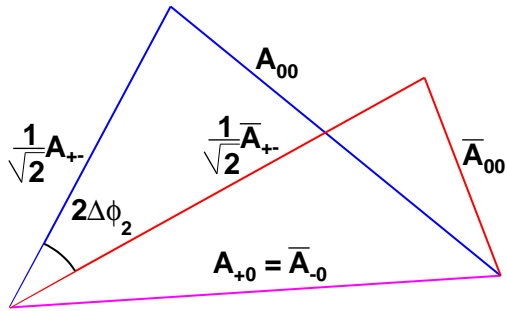} 
  \caption{A sketch of the two isospin triangles, which allow to extract the shift $\Delta\phi_2$.}
  \label{p_phi2}
\end{figure}

All modes presented here are highly background dominated, where especially the continuum ($e^+e^-\to q\bar{q}$, $q=u,d,s$) gives a major contribution. This difficulty is usually overcome by performing a multi-variate analysis or by cutting on event shape dependent variables in order to suppress continuum at the cost of a decrease in the signal detection efficiency.

\section{The Decays \pipi}
\begin{figure}[htb]
\begin{minipage}[]{0.39\columnwidth} 
  \centering
 \includegraphics[height=160pt,width=!]{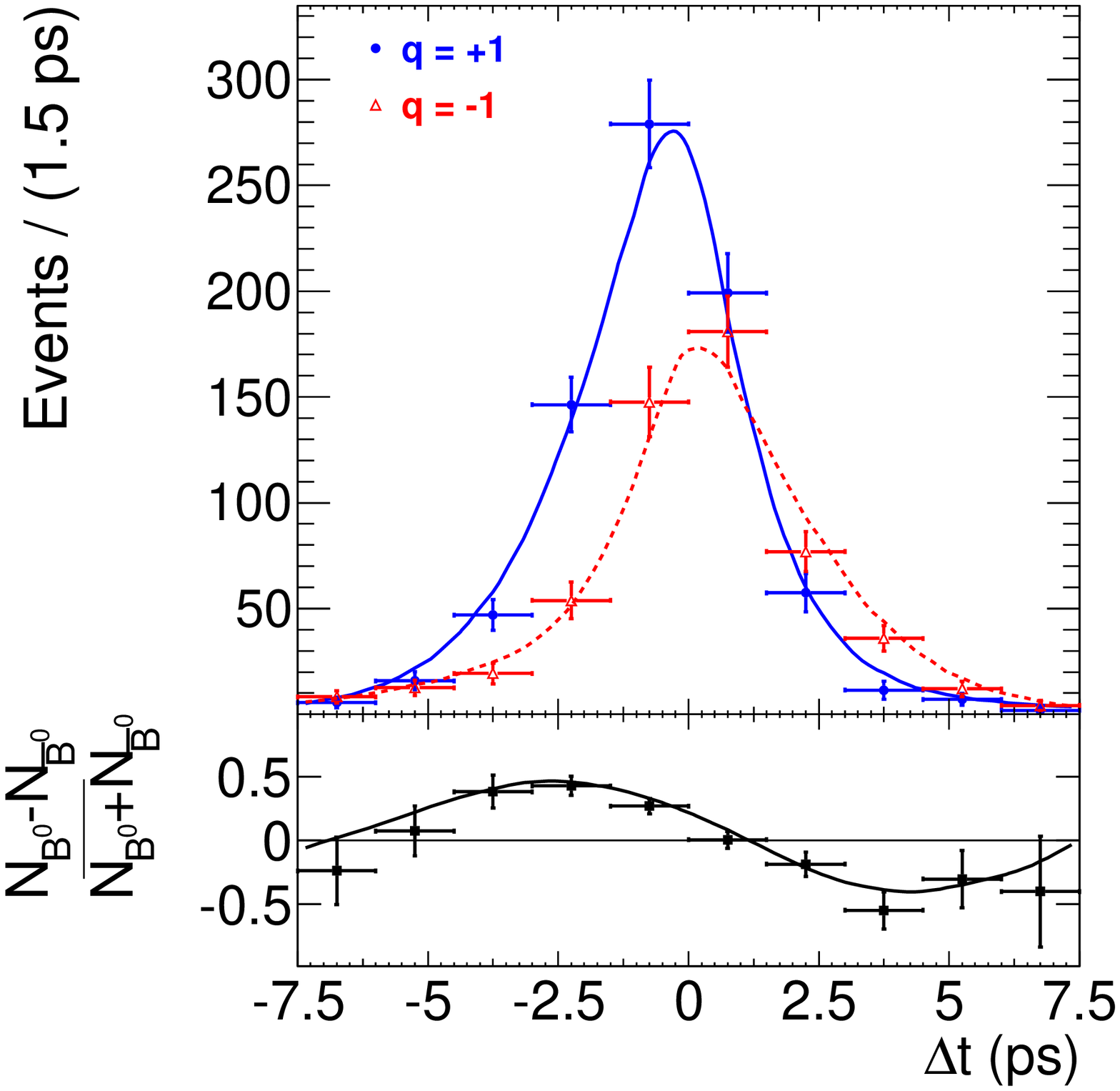} 
\put(-150, 150){a)}
\end{minipage}
\begin{minipage}[]{0.59\columnwidth} 
  \centering
 \includegraphics[height=180pt,width=!]{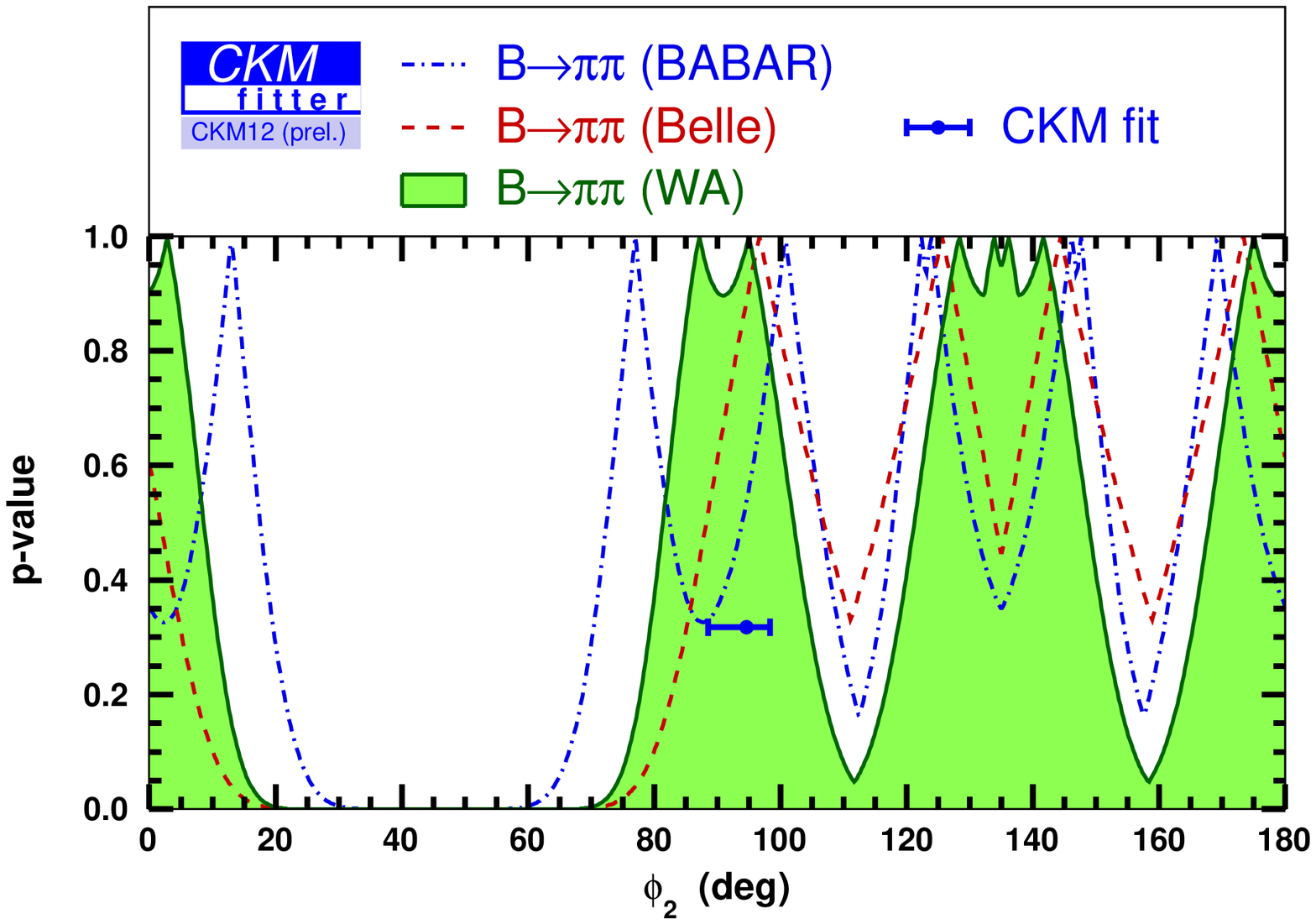} 
\put(-240, 160){b)}
\end{minipage}
  \caption{a) \Dt\ distribution for each flavor tag with the fit result on top and the resulting $CP$ asymmetry from Belle's \pippim\ measurement. Mixing-induced $CP$ violation can be clearly seen in the asymmetry plot. The difference of the integral over the \Dt\ projection indicates direct $CP$ violation. b) $\phi_2$ scan from $B\to\pi\pi$ decays.}
  \label{p_pipi}
\end{figure}

In the decay of the $B$ meson into two pions especially the color-suppressed decay into two $\pi^0$s is experimentally quite difficult: since the $\pi^0$ decays into two photons, making vertex determination only possible if at least one photon converts into a charged lepton pair. The mixing-induced $CP$ asymmetry has not been measured yet. In $B^0\to\pip\pim$ decays both experiments observe mixing-induced $CP$ violation and show a clear presence of penguins, as it can be seen from the height difference of the $\Dt$ distributions for the two flavors of $B_{\rm Tag}$ in Fig.~\ref{p_pipi}. Belle's update to the full data set also includes an improved tracking algorithm and the inclusion of an event shape dependent variable in the fit. Babar's~\cite{pipi_Babar,pipi0_Babar} and Belle's~\cite{pipi_Belle, pipi2_Belle, pi0pi0_Belle, pi+pi0_Belle} results of the measurements of all the individual $B\to\pi\pi$ modes are listed in Table~\ref{t_pipi} and found to be in good agreement. These results are used to obtain an averaged value of $\phi_2 = (87.1^{+17.5}_{-7.8})^{\circ}$~\cite{hfag}. Fig.~\ref{p_pipi} also shows the individual constraints from Babar $\phi_2\in[71^{\circ}, 109^{\circ}]$ and Belle $\phi_2\in[85^{\circ},148^{\circ}]$.

\begin{table}
  \centering
  \begin{tabular}
    {@{\hspace{0.5cm}}l@{\hspace{0.25cm}}| @{\hspace{0.25cm}}l@{\hspace{0.25cm}} | @{\hspace{0.25cm}}l@{\hspace{0.25cm}}}
    \hline \hline
      & BaBar & Belle \\
\hline
\hline
$B^0\to\pip\pip$  & &preliminary \\
\hline
N($B\bar{B}$) &$467$$\times 10^6$ & $772$$\times 10^6$ \\
${\cal B}\times 10^{6}$  &$5.5\pm 0.4 \pm 0.3$ & $5.04\pm0.21\pm0.18 $ \\
${\cal S}_{CP}^{\pip\pim}$ & $-0.68\pm0.10\pm0.03$ & $-0.636\pm0.082\pm0.027$\\
${\cal A}_{CP}^{\pip\pim}$ & $+0.25 \pm0.08\pm0.02$& $+0.328\pm0.061\pm 0.027$\\
\hline
$B\to\pi^0\pi^0$  \\
\hline
 N($B\bar{B}$)  &$467$$\times 10^6$ & $275$$\times 10^6$ \\
${\cal B}\times 10^{6}$ & $1.83\pm0.21\pm0.13$ & $2.3_{−0.5\;-0.3}^{+0.4\;+0.2}$ \\
${\cal A}_{CP}^{\pi^0\pi^0}$ &$+0.43\pm0.26\pm0.05$ &$+0.44^{+0.53}_{-0.52}\pm0.17$ \\
\hline
$B^\pm\to\pi^\pm\pi^0$  \\
\hline
 N($B\bar{B}$) &$383$$\times 10^6$ & $772$$\times 10^6$\\
${\cal B}\times 10^{6}$ &$5.02\pm0.46\pm0.29$ &$5.86\pm0.26\pm0.38$ \\
${\cal A}_{CP}^{\pi^\pm\pi^0}$ &$+0.03\pm0.08\pm0.01 $ & $+0.043\pm0.043\pm0.007$ \\
   \hline
  \end{tabular}
  \caption{Summary of the measurements of $B\to\pi\pi$. The branching fractions of $B^0\to\pip\pim$ are taken from ~\cite{babar_pippimB, belle_pippimB}}
  \label{t_pipi}
\end{table}

\section{The Decays $B\to\rho\rho$}
Since the $\rho$ meson decays into two pions, the $B$ meson has to be reconstructed from a four-pion final state, including a $\pi^0$ in the case of a $\rho^{\pm}$. The presence of other, largely unknown four-pion final states makes the isolation of $B\to\rho\rho$ decays quite challenging. Due to the two vector ($V$) particle final state having contributions from three helicity amplitudes ($A_0,A_+,A_-$), a helicity analysis is needed in order to separate the amplitudes with even and odd $CP$ eigenvalues. In $B$ meson decays into light vector mesons the $CP$ even amplitude $A_0$ is expected to be dominant. However the prediction of color-suppressed $B\to VV$ decays is complicated even at leading order, for example by the non-factorization of the spectator scattering for the transverse amplitude~\cite{BVV_theo}. A measurement of the distribution of the helicity angles between the $\pi$ and the $B$ flight direction in the frame of the $i$-th $\rho$, $\Theta_{Hi}$, is performed and allows measuring the fraction of longitudinal polarization, $f_L = A_0^2/|\sum_{i=0,+,-}A_i^2|$. The decay rate depending on the helicity angles is given by

\begin{equation}
\frac{d\Gamma^2}{d\Gamma(\cos(\Theta_{H1})d\Gamma(\cos(\Theta_{H1})} = \frac{9}{4}(f_L\cos^2(\Theta_{H1})\cos^2(\Theta_{H2}) + \frac{1}{4}(1-f_L)\sin^2(\Theta_{H1})\sin^2(\Theta_{H2})).
\end{equation}   
Including the helicity angles in the fit also improves the separation of the different four-pion final states which otherwise is solely achieved by a fit to the invariant dipion masses. 

Because of the possible presence of many different four-pion final states, interference has to be considered as well. However, since the number of signal events is very small compared with the underlying (total) background, a full amplitude analysis of all the various four pion modes is not feasible with the current data currently available.  Therefore, all the measurements presented here assumed an incoherent model and some accounted for interference effects through a Monte Carlo (MC) study, where samples with maximal constructive or destructive interference between the four-pion final states were generated and then fitted with an incoherent fitter. In all these cases the current statistical error is larger than the uncertainty from neglecting interference, which justifies such a treatment for now. 

\begin{figure}[htb]
\begin{minipage}[]{0.39\columnwidth} 
  \centering
 \includegraphics[height=120pt,width=!]{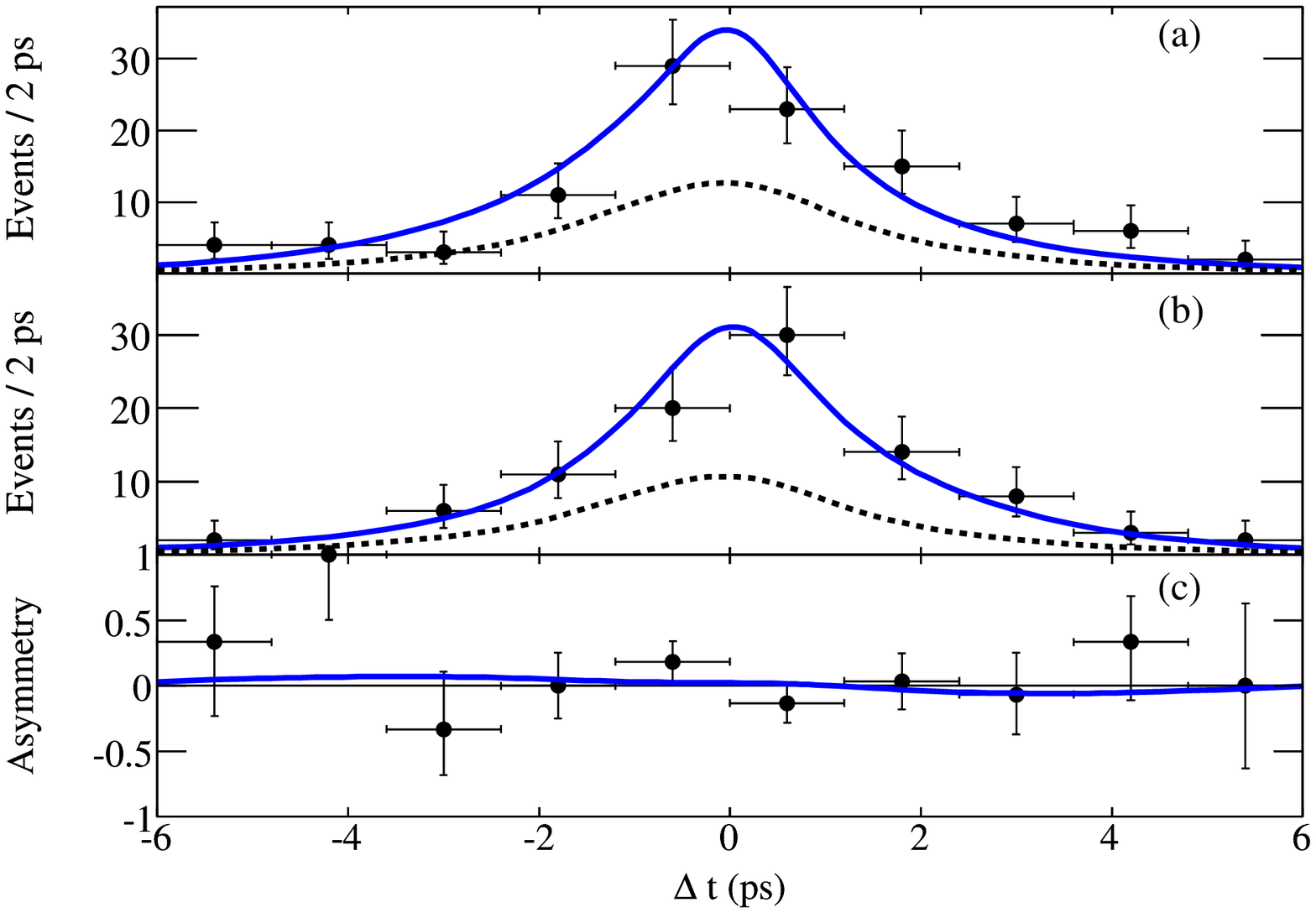} 
\put(-160, 150){i)}
\end{minipage}
\begin{minipage}[]{0.59\columnwidth} 
  \centering
 \includegraphics[height=180pt,width=!]{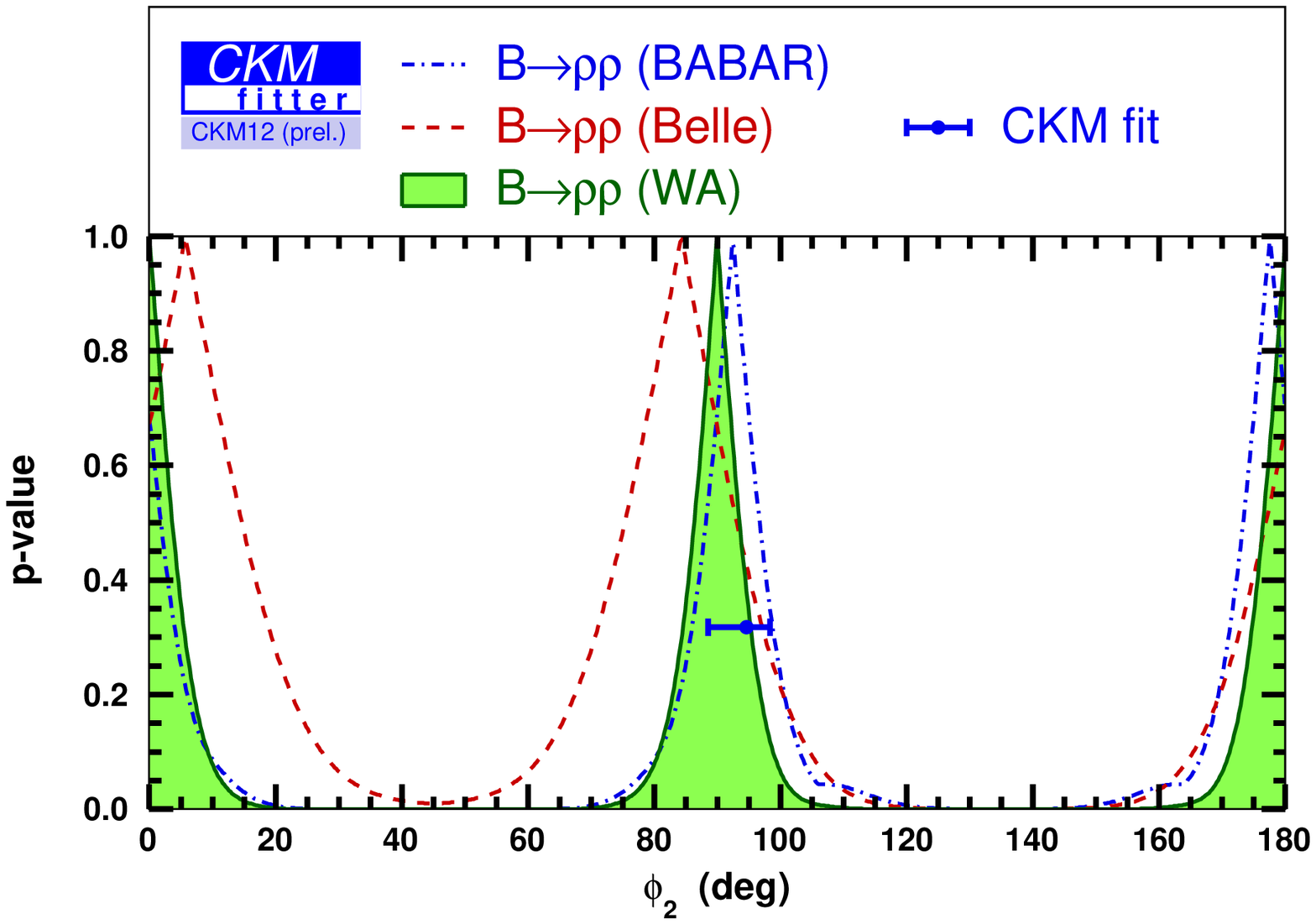} 
\put(-240, 160){ii)}
\end{minipage}
  \caption{i) \Dt\ distribution for each flavor tag with the fit result on top and the resulting $CP$ asymmetry from Babar's $B\to\rho^+\rho^-$ measurement. ii) $\phi_2$ scan from $B\to\rho\rho$ decays.}
  \label{p_rhorho}
\end{figure}

The contribution from penguins is expected to be very small also resulting in a relatively very small decay rate into two $\rho^0$s. This makes the isospin analysis in the $B\to\rho\rho$ system less ambiguous compared to the one in the $B\to\pi\pi$ system: The two isospin triangles are almost degenerate, thus the different solutions overlap. 
Table~\ref{t_rhorho} summarizes the results from Babar~\cite{rhoprhom_BaBar, r0r0Babar, rhoprho0_BaBar} and Belle~\cite{rhoprhom_Belle1, rhoprhom_Belle2, r0r0_Belle, rpr0_belle} which are in good agreement except for the polarization in $B\to\rho^0\rho^0$ decays, where Belle's result is by $2.1\sigma$ lower than Babar's. Fig.~\ref{p_rhorho} shows the $\Dt$ distribution from Babar's $B^0\to\rho^+\rho^-$ measurement with the fit result on top and moreover, the \phitwo\ constraints, where Babar's and Belle's solutions best compatible with the SM are $\phi_2 = (92.4\pm6.4)^{\circ}$ and $\phi_2 =(84\pm13)^{\circ}$, respectively, averaged to $\phi_2 = (89.9^{+5.4}_{-5.3})^{\circ}$.

\begin{table}
  \centering
  \begin{tabular}
    {@{\hspace{0.5cm}}l@{\hspace{0.25cm}}|  @{\hspace{0.25cm}}l@{\hspace{0.25cm}} | @{\hspace{0.25cm}}l@{\hspace{0.25cm}}}
    \hline \hline
      & BaBar & Belle \\
\hline
$B^0\to\rho^+\rho^-$  \\
\hline
N($B\bar{B}$) &$384$$\times 10^6$ & [$275$(${\cal B}$)-$535$($CP$)] $\times 10^6$ \\
${\cal B}\times 10^{6}$  &$25.5\pm2.1^{+3.6}_{-3.9}$ & $22.8\pm3.8^{+2.3}_{-2.6}$\\
$f_L^{+-}$ & $0.992\pm0.024^{+0.026}_{−0.013}$ & $0.941^{+0.034}_{-0.040} \pm 0.030$\\
${\cal S}_{CP}^{\rho^+\rho^-}$ & $-0.17\pm0.20^{+0.05}_{-0.06}$ & $+0.19\pm0.30\pm0.07$\\
${\cal A}_{CP}^{\rho^+\rho^-}$ & $-0.01\pm0.15\pm0.06 $ & $+0.19\pm0.30\pm0.07$\\
\hline
$B\to\rho^0\rho^0$  & & preliminary\\
\hline
 N($B\bar{B}$) &$465$$\times 10^6$  & $772$$\times 10^6$  \\
${\cal B}\times 10^{6}$ & $0.92\pm0.32\pm0.14 $ & $1.02\pm0.30 \pm 0.22<1.5$(UL) \\
$f_L^{00}$ & $0.75^{+0.11}_{-0.14}\pm0.04 $ & $0.21^{+0.18}_{-0.22}\pm0.11$\\
${\cal S}_{CP}^{\rho^0\rho^0}$ &$+0.3\pm0.7\pm0.2$ & n.a. \\
${\cal A}_{CP}^{\rho^0\rho^0}$ &$-0.2\pm0.8\pm0.3$ & n.a. \\
\hline
$B^\pm\to\rho^\pm\rho^0$  \\
\hline
 N($B\bar{B}$) &$465$$\times 10^6$  & $85$$\times 10^6$ \\
${\cal B}\times 10^{6}$ &$23.7\pm1.4\pm1.4$ &$31.7 \pm7.1^{+3.8}_{-6.7}$ \\
$f_L^{\pm0}$ & $0.950 \pm 0.015 \pm 0.006 $  & $ 0.948 \pm 0.106 \pm 0.021$ \\
${\cal A}_{CP}^{\rho^\pm\rho^0}$ &$+0.054 \pm 0.055 \pm 0.010 $ & $0.00\pm0.22\pm0.03$ \\
   \hline
  \end{tabular}
  \caption{Summary of the measurements of $B\to\rho\rho$.}
  \label{t_rhorho}
\end{table}

\section{The Decay \aonepi}
\begin{figure}[htb]
\begin{minipage}[]{0.49\columnwidth} 
  \centering
 \includegraphics[height=!,width=0.8\columnwidth]{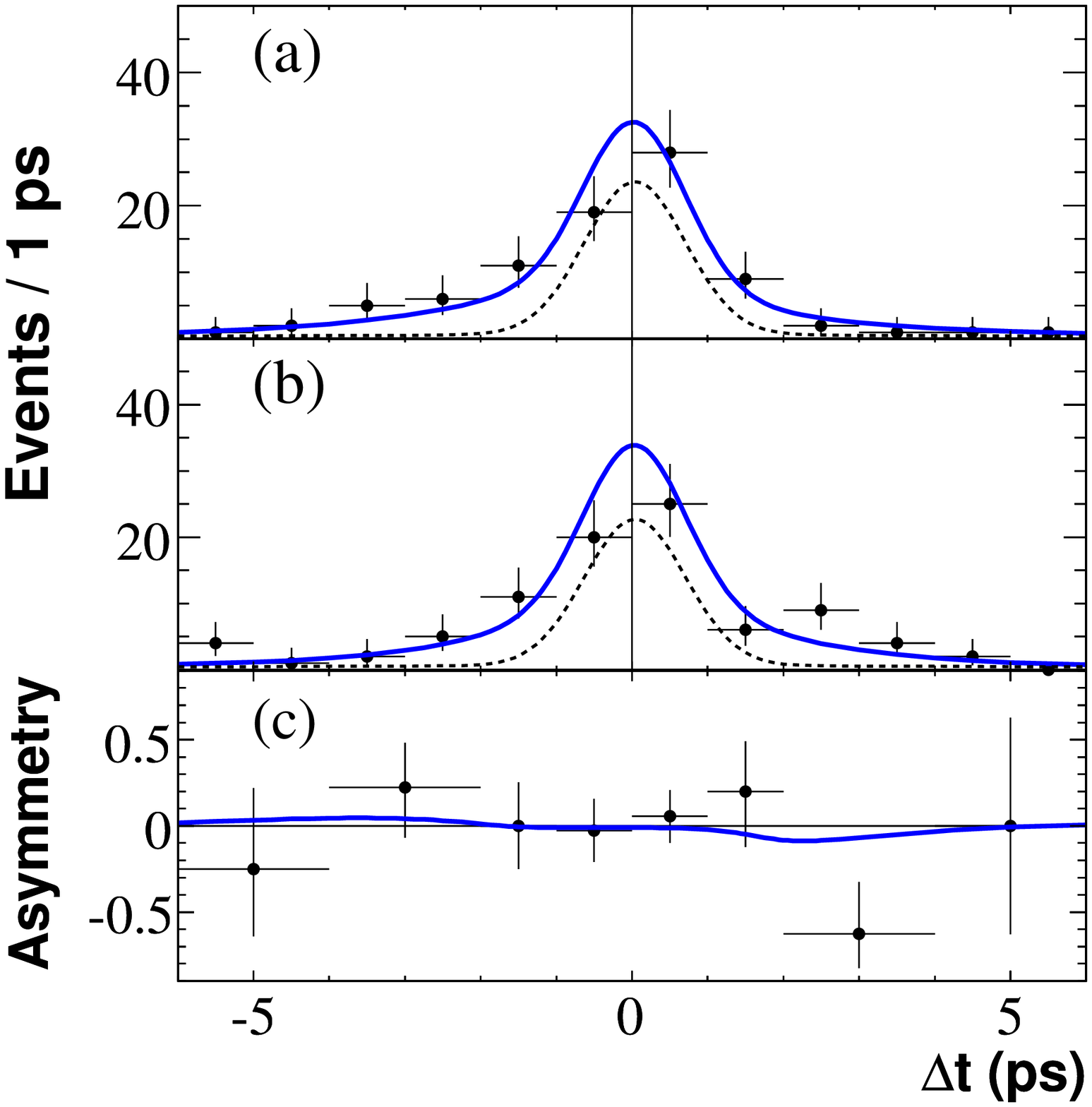}
\end{minipage}
\begin{minipage}[]{0.49\columnwidth} 
  \centering
\includegraphics[height=80pt,width=0.9\columnwidth]{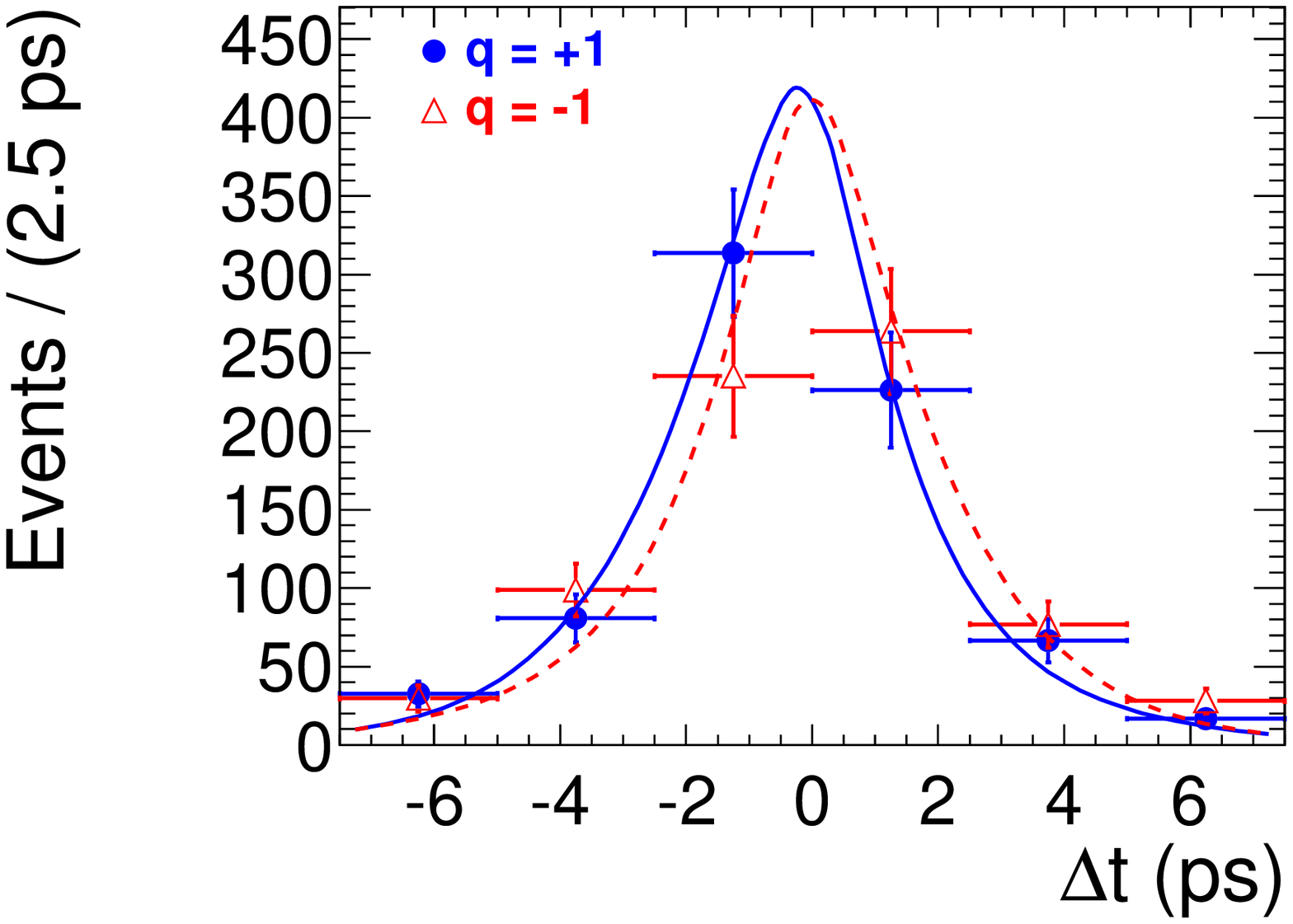} 
\put(-45,60){\includegraphics[height=10pt,width=!]{B-logo.epsf} }\\    
\vspace{-20pt}                                                                                                         
      \includegraphics[height=80pt,width=0.9\columnwidth]{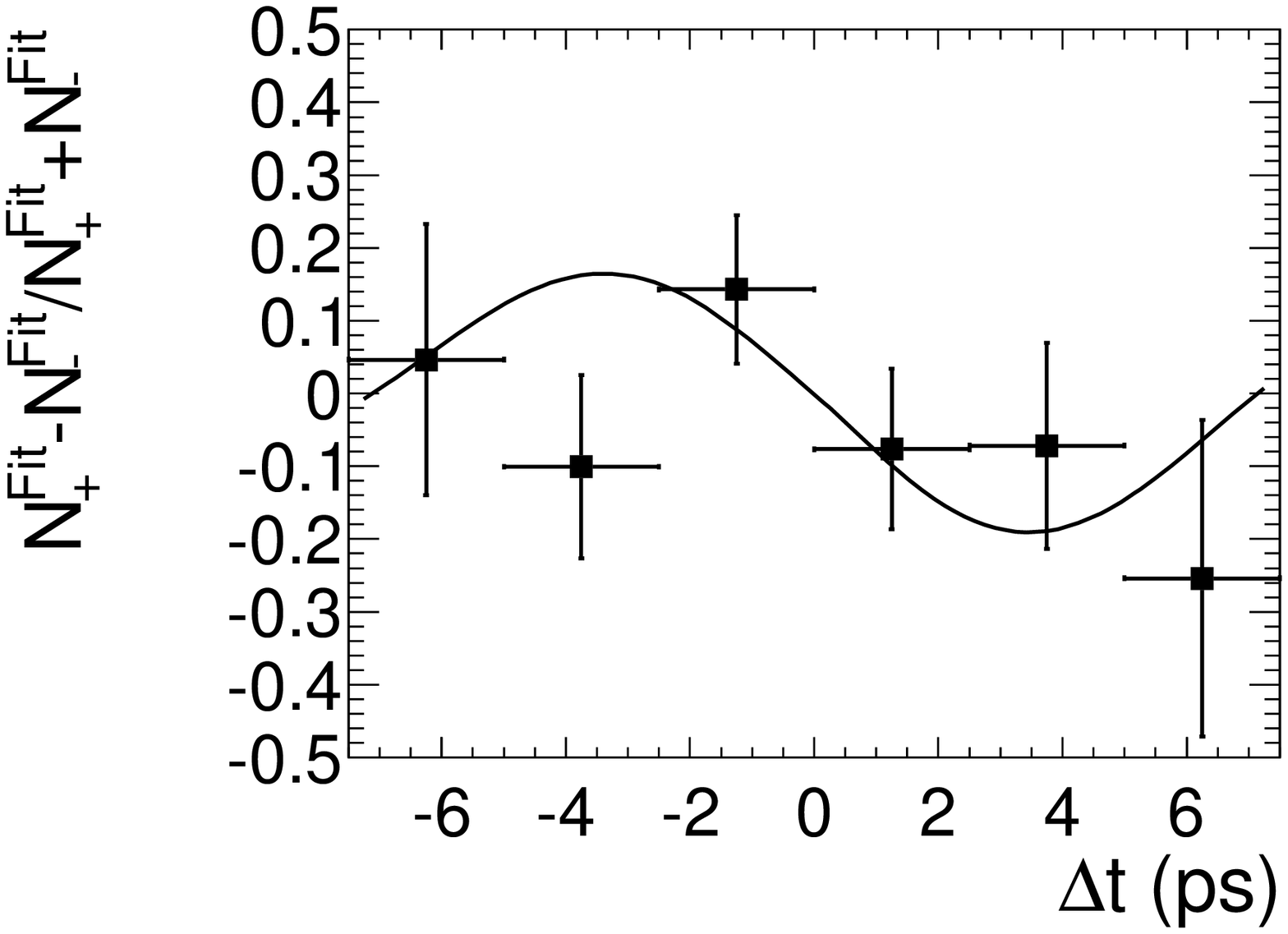}\\                                                    \vspace{-5pt}       
\end{minipage}
  \caption{Projection of the fit result onto $\Delta t$ for \aonepi. Babar's result is shown on the left side. a) $B_{\rm tag}= B$, b) $B_{\rm tag}= \bar{B}$ and c) the resulting time-dependent $CP$ asymmetry. Belle's result is shown on the right side.}
  \label{p_a1pi}
\end{figure}

The decay \aonepi\ provides another four charged pion final state sensitive to \phitwo, but this time a decay into a non-$CP$ eigenstate. Therefore Eq.~\ref{eq1} has to be extended to five $CP$ parameters that in addition allow to measure the time and flavor integrated charge asymmetry, the rate asymmetry between $a_1$ being formed from the spectator quark or not and a strong phase difference between the different amplitudes involved~\cite{nonCPeigenstates, nonCPeigenstates2}. Belle reported first evidence of mixing induced $CP$ violation in this mode with a sigificance of $3.1\sigma$; $S_{CP} = -0.51 \pm 0.14 \;(\rm stat) \pm \;(\rm 0.08)$~\cite{a1pi_Belle} while Babar measured ${\cal S}_{CP} = +0.37 \pm 0.21 (\rm stat) \pm 0.07 (\rm syst)$~\cite{a1pi_Babar}. Although there is a discrepancy in the sign of ${\cal S}_{CP}$, the asymmetry plot shown below the $\Delta t$ distributions in Fig.~\ref{p_a1pi}, agree. The amount of penguin pollution can in general be estimated using $SU(3)$ symmetry~\cite{CKMangles} but would need more input data. 

\section{The Decay $B\to(\rho\pi)^0$}

\begin{figure}[htb]
\begin{minipage}[]{0.49\columnwidth} 
  \centering
 \includegraphics[height=!,width=0.95\columnwidth]{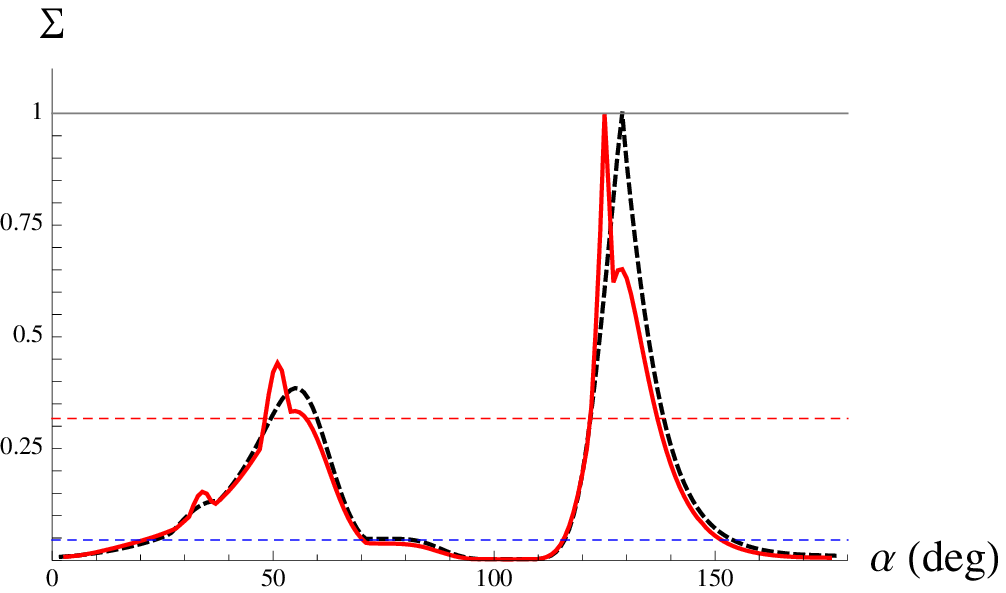}
\put(-210, 100){i)}
\end{minipage}
\begin{minipage}[]{0.49\columnwidth} 
 \includegraphics[height=!,width=0.95\columnwidth]{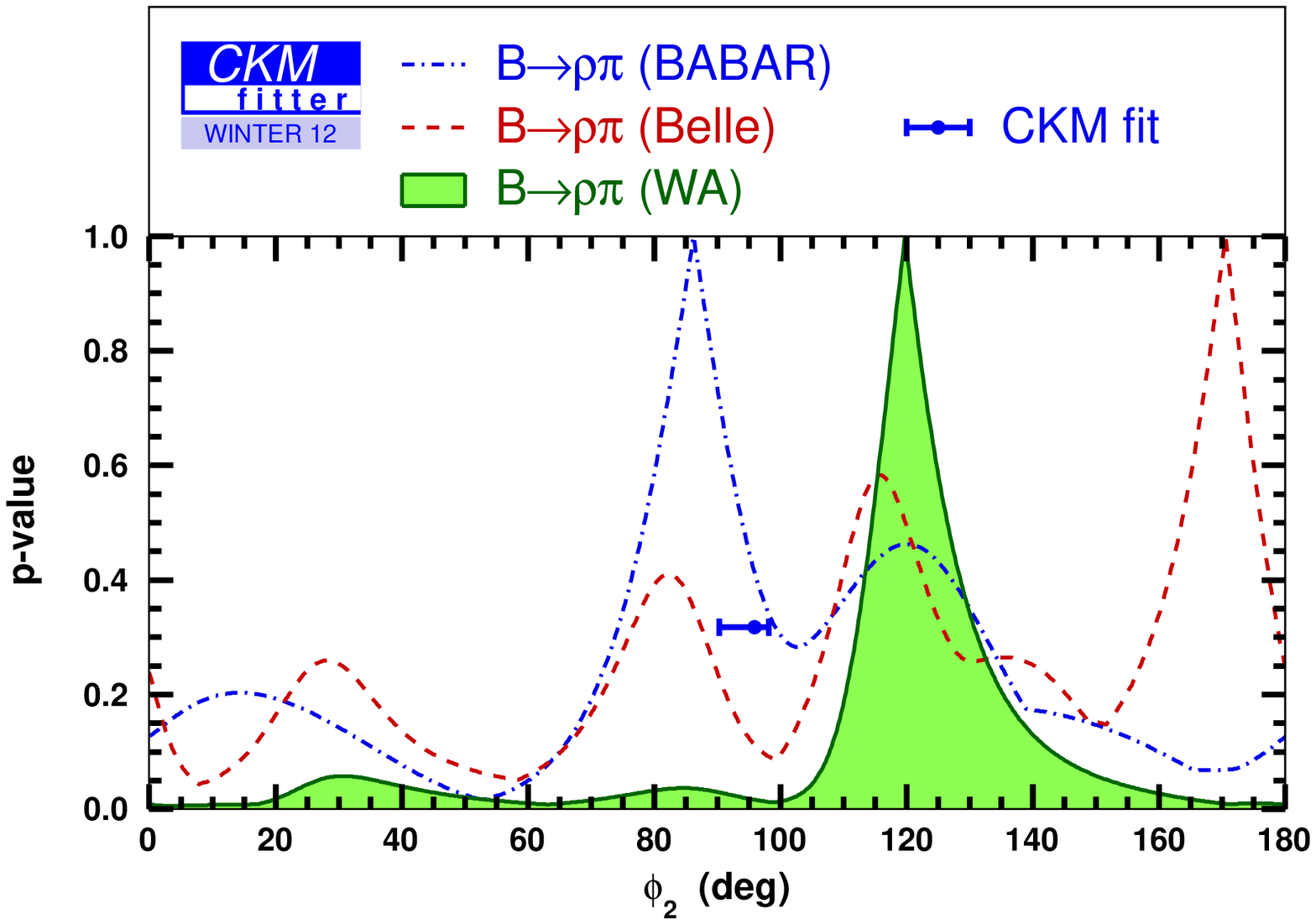}
\put(-200, 110){ii)}
\end{minipage}
  \caption{i) $\phi_2$ scan from the Babar update. ii) a scan of both, the previous Babar and Belle's, measurements combined. } 
  \label{p_rhopi}
\end{figure}

Although, these decays include also non-$CP$ eigen states in the final state that are, it is possible to obtain a single solution for $\phi_2$ via a time-dependent amplitude analysis. The quasi-two body approach is replaced by the $UV$ formalism~\cite{UV}, where the complex decay amplitudes are constructed from isospin relations and described by 27 real parameters which can be related to the quasi two-body $CP$ asymmetries. This allows measuring the phase differences between the interfering $\rho$ resonances from $B\to(\rho\pi)^0$ decays in a dalitz plot and hereby access $\phi_2$ directly without any ambiguity. Belle~\cite{rhopi_belle} performed this analysis on $449\times 10^6$ $B\bar{B}$ pairs and constrained $\phi_2\in[68^{\circ},95^{\circ}]$ from a probability scan where a large SM-disfavored region remains. Babar~\cite{rhopi_babar1, rhopi_babar2} updated their previous analysis to $471\times 10^6$ $B\bar{B}$ pairs where also particle identification, charged particle tracking and a multivariate discriminator, which is also used as a fit variable, was improved. The $CP$ parameters obtained, shown in Table~\ref{t_rhopi}, are in good agreement. However, a robustness study performed by Babar demonstrated that, different to the $CP$ and $UV$ parameters, the $\phi_2$ scan is not robust with the statistics currently available and hence cannot be interpreted in terms of Gaussian statistics. Fig~\ref{p_rhopi} shows the $\phi_2$ scan from the Babar update and a scan of both, the previous Babar and Belle's, measurements combined.

\begin{table}
  \centering
  \begin{tabular}
    {@{\hspace{0.5cm}}l@{\hspace{0.25cm}}| @{\hspace{0.25cm}}l@{\hspace{0.25cm}} | @{\hspace{0.25cm}}l@{\hspace{0.25cm}}}
    \hline \hline
      & BaBar & Belle \\
\hline
\hline
$B^0\to(\rho\pi)^0$  & preliminary& \\
\hline
N($B\bar{B}$) &$471$$\times 10^6$ & $449$$\times 10^6$ \\
${\cal S}_{CP}^{\rho^{\pm}\pi^{\mp}}$ & $0.053 \pm 0.081 \pm 0.034$ & $0.06 \pm 0.13 \pm 0.05 $\\
${\cal A}_{CP}^{\rho^{\pm}\pi^{\mp}}$ & $−0.100 \pm 0.029 \pm 0.021$& $-0.12 \pm 0.05 \pm 0.04$\\
${\cal C}_{CP}^{\rho^{\pm}\pi^{\mp}}$ & $0.016 \pm 0.059 \pm 0.036$& $-0.13 \pm 0.09 \pm 0.05$\\

${\cal S}_{CP}^{\rho^{0}\pi^{0}}$ & $−0.37 \pm 0.34 \pm 0.20$ & $0.17 \pm 0.57\pm 0.35$\\
${\cal C}_{CP}^{\rho^{0}\pi^{0}}$ & $0.19 \pm 0.23 \pm 0.15$& $0.49 \pm 0.36 \pm 0.28$\\
   \hline
  \end{tabular}
  \caption{Summary of the measurements of $B\to(\rho\pi)^0$. Since final state that are no $CP$ eigen states are involved, ${\cal A}_{CP}$ is the time and flavor integrated charge asymmetry, while ${\cal C}_{CP}$ quantifies the flavor-dependent direct $CP$ violation.}
  \label{t_rhopi}
\end{table}

\section{Summary}
We have presented a summary of $\phi_2$ related measurements from Belle and Babar. The modes presented here include $B\to\pi\pi$, $B\to\rho\rho$, $B^0\to a_{1}^{\pm}\pi^{\mp}$ and $B^0\to(\rho\pi)^0$. While most channels have been measured using the full data sample, some updates are still anticipated.

The world averages of \phitwo\ as computed by the CKMfitter~\cite{CKMfitter} (including the latest $B\to\pi\pi,\rho\rho$ results) and UTfit~\cite{UTfit} collaborations are  $\phitwo = (88.5^{+4.7}_{-4.4})^{\circ}$ and $\phitwo = (88.7 \pm 3.1 )^{\circ}$, respectively, the probability scans are shown in Fig.~\ref{p_phi2}. 

It was proposed to use the new measurement of $B\to\pi\pi$ in order to update the constraint on the CKM angle $\phi_3 \; (\gamma)$, as described in~\cite{phi3}. 

So far, no tension with the very successful SM has been seen. Nevertheless, with BelleII being built and LHCb operating, the next generation of $B$ physics experiments are expected to further reduce the uncertainty of the CKM observables: The uncertainty of \phitwo, for example, is expected to be reduced to $1^{\circ}-2^{\circ}$~\cite{Belle2}. 
\begin{figure}[htb]
\begin{minipage}[]{0.49\columnwidth} 
  \centering
 \includegraphics[height=!,width=0.95\columnwidth]{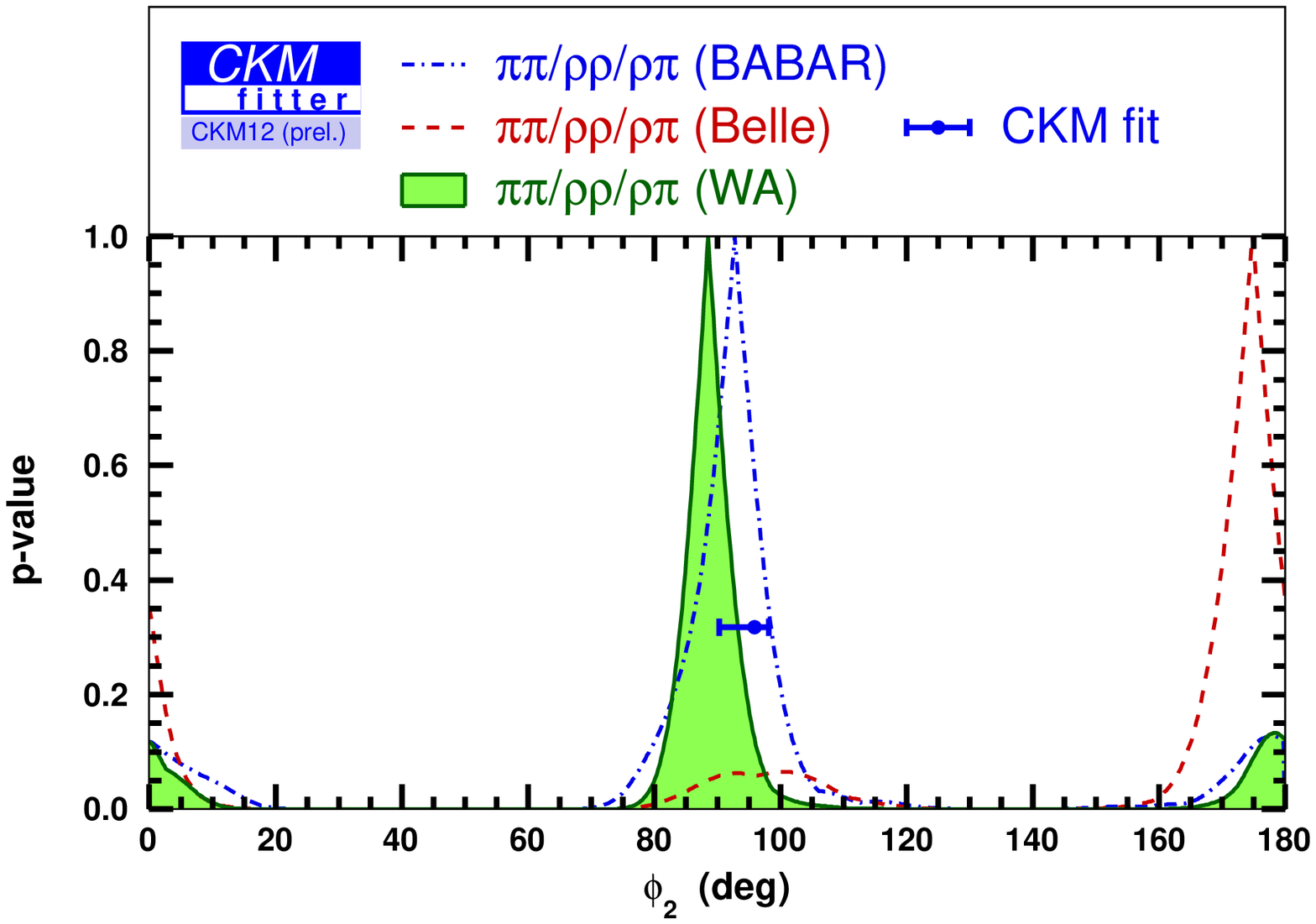}
\end{minipage}
\begin{minipage}[]{0.49\columnwidth} 
 \includegraphics[height=!,width=0.95\columnwidth]{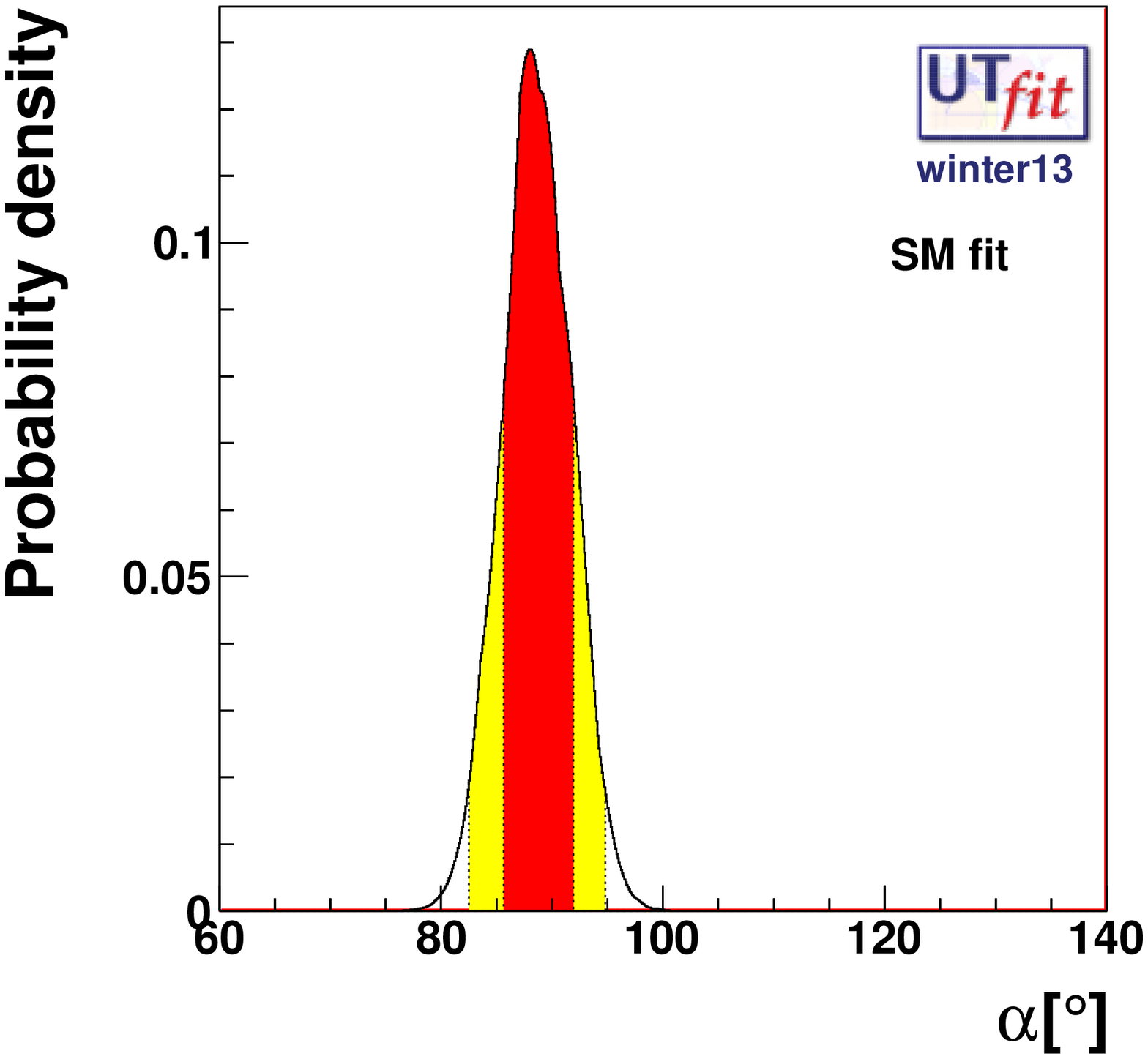}
\end{minipage}
\caption{$\phi_2$ scan with $B\to \pi\pi, \rho\pi, \rho\rho$ combined from the CKM fitter and the UTfit collaborations. } 
 \label{p_phi2}
\end{figure}

\end{document}